\documentclass[aps,prl,twocolumn,longbibliography,superscriptaddress]{revtex4-1}
\usepackage{graphicx} 
\usepackage{epstopdf}
\usepackage{amsmath}
\usepackage{color}
\usepackage{multirow}
\usepackage{amssymb}
\usepackage{dcolumn}%
\usepackage{bm}%
\usepackage{ctable}
\usepackage{pbox}
\usepackage{tabularx}
\usepackage{graphicx}
\usepackage[normalem]{ulem}

\newcommand{\SK}[1]{\textcolor{black}{{#1}}}

\begin{document}

\begin{titlepage}

%\title[]{\Large \bf Stability of memory in amorphous solids}
\title{Encoding Robust and Fast Memories in Bulk and Nanoscale Amorphous Solids}

\author{Monoj Adhikari}
\thanks{Equal contributions}
\affiliation{Tata Institute of Fundamental Research, 36/P, Gopanpally Village, Serilingampally Mandal, Ranga Reddy District, Hyderabad 500046, Telangana, India}
\author{Rishabh Sharma}
\thanks{Equal contributions}
\affiliation{Tata Institute of Fundamental Research, 36/P, Gopanpally Village, Serilingampally Mandal, Ranga Reddy District, Hyderabad 500046, Telangana, India}
\author{Smarajit Karmakar}
\email[Corresponding author: ]{smarajit@tifrh.res.in}
\affiliation{Tata Institute of Fundamental Research, 36/P, Gopanpally Village, Serilingampally Mandal, Ranga Reddy District, Hyderabad 500046, Telangana, India}

\begin{abstract}
\SK{We investigate the memory effects under oscillatory shear deformation of amorphous solids through computer simulations. Applications of shear deformations in all orthogonal directions show that encoded memories via this protocol are more robust while performing reading. Our extensive system size analysis of memory effects shows that memory encoding in small systems is faster than in larger systems and is probably impossible in thermodynamically large system sizes. In addition to demonstrating how to encode robust memories in 3D bulk amorphous materials, we devise protocols for encoding and reading memories in pseudo-1D materials in the form of amorphous nano-rods. With this, we show that memory encoding and retrieving can also be done in systems with open surfaces, which all materials would necessarily have in practice, and is thus essential to capitalise on the effectiveness of smaller system sizes to encode memories faster. All in all, we provide protocols for encoding robust and faster memories in amorphous solids both at bulk and nanoscale.} 
\end{abstract}

\maketitle

\end{titlepage}

\noindent{\bf Introduction:}
\SK{Memory is a ubiquitous concept that can be seen both in biological and mechanical systems and is often thought of as a defining feature of non-equilibrium systems \cite{keim2019memory}. Although the existence of history dependence/memory is well-established, the challenge usually is to bring out this aspect by quantifying it using some well-stated protocols. Classic instances of memory in materials might range from shape memory in alloys \cite{Kbhattacharya} or hysteretic memory in magnetic materials \cite{pierceetal}. A simple model of Non-Brownian suspension \cite{natphysCorte,naturePine} was seen to retain the memory of the amplitude of the oscillating drive \cite{PhysRevLett.107.010603,PhysRevLettPaulsen}. Similar memory was also found in the much more complex case of glasses. In these systems, memory might take the form of the system ``remembering" the mechanical drive (the amplitude of the oscillations applied) \cite{MemFiocco,adhikari2018memory} or the thermal perturbations that it was subjected to. The case of ageing and rejuvenation is another interesting aspect where the material seems to remember the temperature at which it was aged \cite{scalliet2019rejuvenation}. The nature of memory that a system can be made to store, along with key attributes like the speed of encoding and reading, the possibility of storing multiple memories, its persistence, etc., all offer valuable insights towards understanding the underlying landscape. 
%For instance, the ability of structural glasses to store multiple memories persistently differentiates it from non-Brownian suspensions that can store only a single memory, both under the same kind of driving protocols. This points to the inherent differences between the energy landscapes of these two types of disordered systems.
In this article, we focus on two crucial aspects of memory in glasses: the speed of reading and writing and fault tolerance. In particular, we provide ways that can be used to encode and read memories faster and in a robust manner. 
}

\SK{A wide range of disorder systems have been found to have the ability to store and read the memory \cite{SethnaPRL, CDWLittlewood, PhysRevLetMiddleton, PhysRevLettPaulsen,PhysRevLettLilly, PhysRevBGilbert,bandi2017d, PhysRevLettLahini, keim2020global,lindeman2021multiple,chattopadhyay2022inter}. Specifically, there has been a notable interest in amorphous solids subjected to cyclic shear deformation. The amorphous solids exhibit a non-equilibrium phase transition under the application of cyclic shear deformation \cite{PREFiocco, regev2013reversibility, Priezjev2013c, PKetal, bhaumik2021role, yeh2020glass, bhaumik2022avalanches, adhikari2022yielding, ozawa2023creating, mutneja2023yielding}.
When the deformation amplitude is smaller than a critical value known as the yielding amplitude ($\gamma_Y$), the system reaches an absorbing state where the configurations do not change when looked at stroboscopically (at the end of the cycle at zero deformation $\gamma=0$). Conversely, when the deformation amplitude exceeds $\gamma_Y$, the configurations change, and the system reaches a diffusive state. In the absorbing phase, following a transient period, the system converges to a limit cycle, cycling through the same sequence of states, effectively retaining a `memory' of the deformation amplitude \cite{MemFiocco,adhikari2018memory,mukherji2019strength,keim2020global,schwen2020embedding,arceri2021marginal,benson2021memory}.
Upon analyzing these states, Ref.\cite{mungan2019networks} showed that the memory observed in glasses exhibits partial Return Point Memory (RPM). These encoded memories exhibit robustness when subjected to various reading protocols \cite{adhikari2018memory}. Furthermore, a recent study in Ref.\cite{shohat2023dissipation} suggests that the microscopic signals associated with this form of memory can manifest in macroscopic quantities such as energy dissipation. These memory signals can be enhanced by introducing asymmetric shear deformation protocol during training and reading, as was observed in a different model system studied in  Ref.\cite{jalowiec2023isolating}.}

\SK{However, one common theme that emerges in all these protocols is that the correct reading of the memory relies on the underlying assumption that the shear direction during reading and training is the same. This points towards the possible practical limitations of storing memory in these solids. In this work, we want to investigate how to circumvent some of these issues by employing multiple shear directions during memory encoding. We would also like to understand the effects of system size on memory formation, especially the time taken to encode the memory in the sample, and whether large systems or smaller systems are better for encoding memory. We show via extensive computer simulations that encoding memory using a multi-directional oscillatory shear protocol is a much more robust method. This is because it does not require any prior knowledge to read the information. We also see that smaller systems are more suitable for practical reasons associated with the time taken for encoding, as encoding time seems to diverge with system size as a power law. Furthermore, we show that encoding and reading memory in the presence of open surfaces is possible using amorphous nano-rods, thus going from three-dimensional bulk storage to pseudo one-dimensional storage devices.}   

\vskip +0.1in
\noindent{\bf Simulation details:}
\SK{We have studied a well-known glass-forming model, the Kob-Andersen Binary mixture ($A_{80}B_{20}$)  with Lennard Jones interactions between particles (BMLJ) \cite{PhysRevKob}. The details of the interaction potential is given in the SM. 
We simulate BMLJ samples consisting of $N = 500-50000$ particles in 3 dimensions while keeping the number density fixed, $\rho = 1.2$ ($\rho=N/V$, $V$ is the volume of the simulation box). Initially, the system is equilibrated at a reduced temperature $T=1.0$ following a constant temperature molecular dynamics simulation(NVT). After equilibration, the systems are subjected to energy minimization using the conjugate-gradient algorithm to obtain inherent structures \cite{sastry1998signatures}. We use the Athermal Quasi-Static (AQS) procedure \cite{maloney2004subextensive,PREFiocco} to perform cyclic shear deformation as described below. Samples undergo oscillatory shear deformation, where a single cycle is defined as follows 0 $\rightarrow$ $\gamma_{max}$ $\rightarrow  0 \rightarrow -\gamma_{max} \rightarrow 0$, with $\gamma_{max}$ representing the deformation amplitude. A large number of cycles are applied repeatedly until they reach a steady state. This phase is referred to as training. After training, we perform a reading operation which implies performing a single cycle of shear deformation with varying amplitude: $0$  $\rightarrow$ $\gamma_{read}$ $\rightarrow  0 \rightarrow -\gamma_{read} \rightarrow 0$. The simulations for bulk systems with periodic boundary conditions (PBC) reported here are performed in LAMMPS \cite{plimpton1995fast}. The simulations of nanorods are done using our in-house parallel molecular dynamics code. For the reading protocol, our measured quantity is the mean squared displacement (MSD). We denote $MSD_o$ as the mean squared displacement when measured with respect to the original configuration. Further details of the simulation protocol can be found in the SM.} 

\begin{figure}[htp]
%\centering
\includegraphics[width=0.49\textwidth]{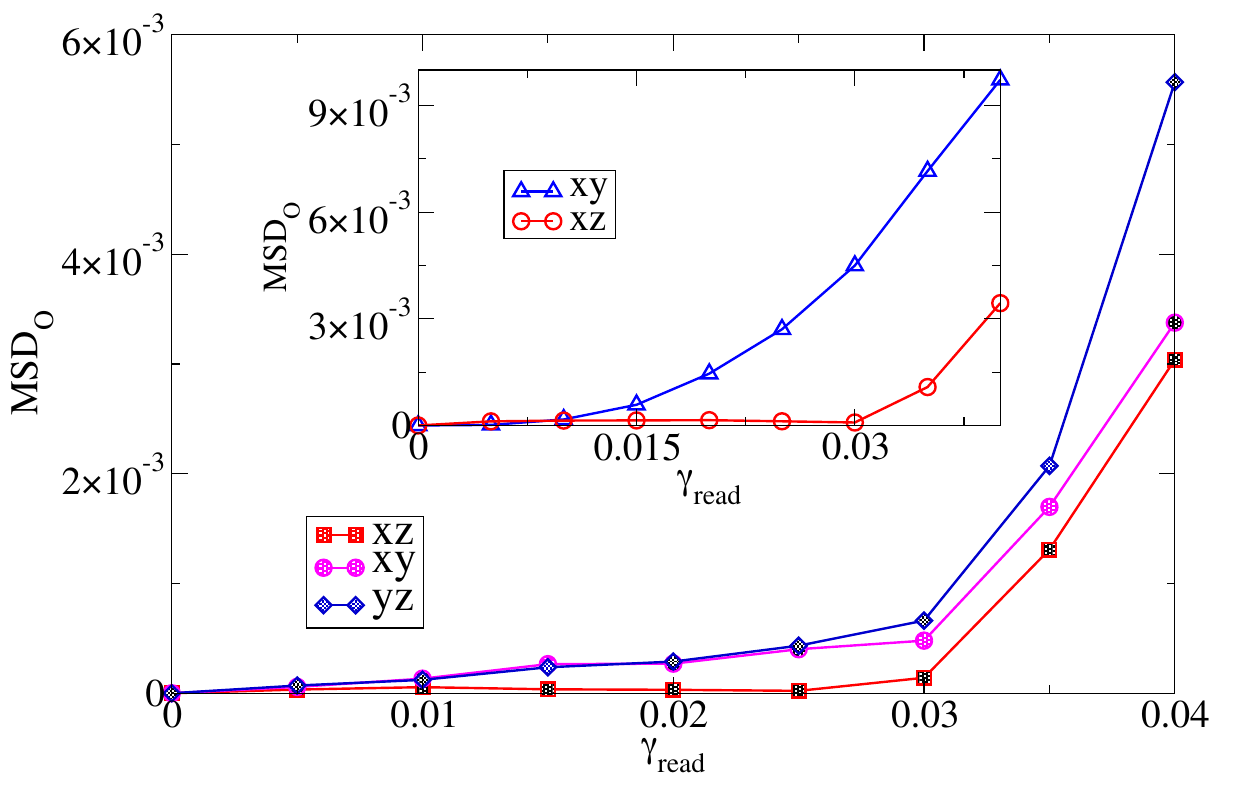}
\caption{$MSD_o$ as a function of $\gamma_{read}$. Inset: The system is trained at $\gamma_{max}=0.03$ with the shear direction being $xz$ for two different reading protocols where the shear direction is varied. When the direction of shear during reading and training is the same, the MSD shows a kink at the training amplitude, whereas there is no kink when the direction of shear during reading and training is different. Main Panel: $MSD_o$ as a function of $\gamma_{read}$ for the system trained with three different shear directions: $xz$, $xy$, $yz$. The starting configuration for the reading operation is the sample where the $xz$ direction of shear is applied last. The training amplitude is $0.03$. For all reading protocols, $MSD_o$ shows a kink at the encoded training amplitude irrespective of the shear directions.}
\label{fig.olddir_mem}
\end{figure}

\vskip +0.05in
\noindent{\bf Results:}
\SK{We investigate the stability of memory under the different shear directions. We have taken a system that is trained at $\gamma_{max} = 0.03$ with the shear direction being $xz$. Previous studies \cite{MemFiocco,adhikari2018memory} show when the direction of shear during training and reading is the same, a kink is observed at the training amplitude. We now ask what happens if the shear direction during reading differs from the shear direction during training. In the inset of Fig. \ref{fig.olddir_mem}, we show $MSD_o$ as a function of $\gamma_{read}$. When the shear direction of training and reading is the same, which is $xz$ direction, as expected, $MSD_o$ shows a kink. However, no kink is observed if the shear direction is chosen differently than the training. In this case, we choose $xy$. This result signifies that the memory encoded with fixed amplitude and a single direction of shear deformation is not robust under perturbations and possible reading mistakes; one needs to know the shear direction of training beforehand to read the memory correctly and indestructibly. The encoded memory is permanently lost if the reading is mistakenly done in the wrong shear direction. We want to explore a training protocol where memory could be stable in any shear direction and thus be robust during the reading protocol.}  

\begin{figure*}[htpb]
\centering
\vskip -0.1in
\includegraphics[width= 0.99\textwidth]{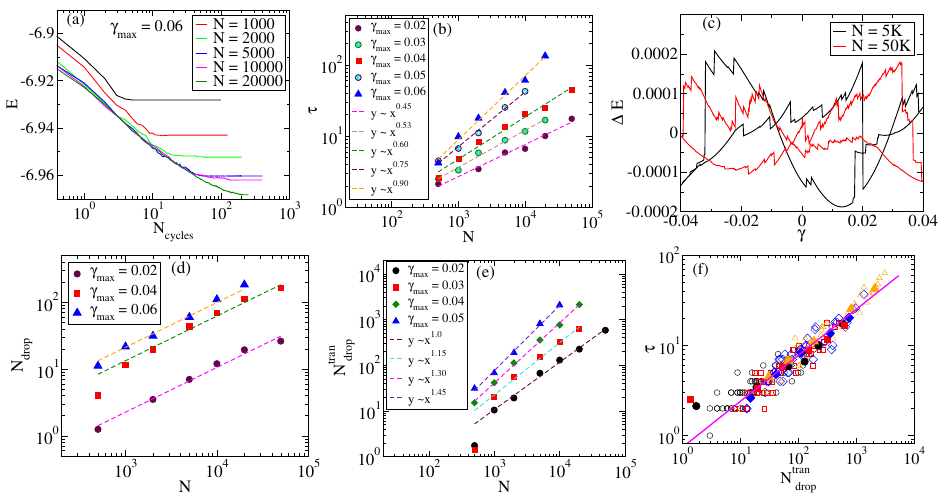}
%\includegraphics[width= 0.48\textwidth]{MultiDir/EvsCycle_gamma06.eps}
%\includegraphics[width= 0.47\textwidth]{MultiDir/tauvs_N_allgamma.eps}
%\vskip -0.3in
\caption{(a) The system is trained with $\gamma_{max} = 0.06$. Stroboscopic energy, $E(\gamma=0)$ is plotted as a function of $N_{cyclces}$. The steady state for different system sizes reaches at an increasingly larger number of cycles with increasing system sizes. (b) The relaxation time to reach the steady state is plotted as a function of system sizes. We observe that $\tau \sim N^{\beta}$, where $\beta$ is different for different $\gamma_{max}$. (c) $\Delta E$ is plotted as a function of $\gamma$ in the steady state for two different system sizes, as shown in the legend. $\Delta E = E_{\gamma}-E_{fit}$, where $E_{\gamma}$ is the actual energy in the steady state in a cycle and $E_{fit}$ is obtained by parabolic fitting of the $E_{\gamma}$. (d) The number of drops is plotted as a function of system sizes at the steady state. A number of drops also show a power law, and the slope is the same for every $\gamma$. (e) Number of drops ($N^{tran}_{drop}$) to reach the steady state as a function of system size, $N$ for different $\gamma_{max}$. (f) $\tau$ is plotted as a function of total number drops to reach the steady state for all $\gamma$ for different $N$. The line has a slope of around $0.5$.}
\label{fig.PBC_Sys}
\end{figure*}
\SK{Recently, it has been found that a system reaches a steady state with lower energy when shear is applied in all orthogonal directions of shear, $xy$, $xz$ and $yz$ \cite{krishnan2023annealing} for a three-dimensional system. We want to understand whether the steady state obtained via oscillatory shear applied on all the orthogonal shear directions sequentially encodes memory; if the answer is yes, then how stable those memories are. We have taken a sample that is trained at an amplitude of $0.03$. First, we apply an oscillatory shear cycle in the $xy$ plane, then $xz$ and $yz$. We repeat this sequence for a large number of cycles until the absorbing state is reached. In this protocol, we could use three different stroboscopic configurations (stroboscopic under any of the three orthogonal directions) for reading. We have taken a configuration where the $xz$ direction of shear is applied last. Then, we perform reading operations with varying directions of shear. When the direction of shear during reading is $xz$, the expected kink is observed at $\gamma_{read}=0.03$.
Interestingly, $MSD_o$ shows a kink even when the shear direction during reading is $xy$ or $yz$ as shown in Fig. \ref{fig.olddir_mem} (main panel). Similar behaviour can be observed for any starting configuration. Thus, training with multi-direction of shear is more stable for encoding any memory. We need not worry about the shear direction of training while performing reading.} 

\SK{Next, we want to systematically investigate the effect of system size on memory formation in amorphous solids to ascertain the suitability both in terms of stability and operational difficulty of memory encoding under different system sizes. We prepare systems of different sizes ranging from $500$ to $50000$. We apply shear deformation with fixed amplitude in each system for many cycles. In Fig. \ref{fig.PBC_Sys} (a), we show stroboscopic energy as a function of the number of cycles for different system sizes. We observe that with increasing system sizes, the number of cycles to reach the steady state is increasing. The final energy is also becoming less and less with increasing system sizes. When we plot the number of cycles to reach the steady state, $\tau$, we observe it grows with system size as a power law with an exponent $\beta$ as $\tau \sim N^{\beta}$, with $\beta$ changing from $0.45$ for small amplitude of $\gamma_{max} = 0.02$ to $0.90$ for $\gamma_{max} = 0.06$ systematically as shown in Fig.\ref{fig.PBC_Sys}(b). Note that the estimated yield strain for this system is $\gamma_Y \simeq 0.07$. Thus, this observation implies that the system will never find a steady state for an infinitely large system. Correspondingly, it will be difficult to store memory in a large system size than when the system size is small.}  

\SK{To understand the reason for the increase in the number of cycles $\tau$ with system size, we look at the number of plastic events the system encounters during the oscillatory shear cycles.  It is well known that the number of plastic events increases with increasing system size \cite{karmakar2010statistical}. 
Under oscillatory shear deformation, the stroboscopic configurations become the same when the system reaches a steady state. However, between the cycles, the system jumps from one minimum to another, and correspondingly, many plastic drops can be observed in a cycle. 
%\begin{figure}[htpb]
%\centering
%\includegraphics[width= 0.49\textwidth]{MultiDir/StabilityMemory_Fig3New.pdf}
%\includegraphics[width= 0.48\textwidth]{MultiDir/Compare_N5vs50k.eps}
%\includegraphics[width=0.47\textwidth]{MultiDir/trans_dropvsN_allgamma_corrected}
%\includegraphics[width=0.48\textwidth]{MultiDir/NumberofDrop_N_allgamma.eps}
%\includegraphics[width= 0.48\textwidth]{MultiDir/gamma06tauvsN.eps}
%\caption{Top: $\Delta E$ is plotted as a function of $\gamma$ in the steady state for two different system sizes as shown in the legend. $\Delta E = E_{\gamma}-E_{fit}$, where $E_{\gamma}$ is the actual energy in the steady state in a cycle and $E_{fit}$ is obtained by parabolic fitting of the $E_{\gamma}$. Bottom: Number of drops ($N^{tran}_{drop}$) to reach the steady state as a function of system size, $N$ for different $\gamma_{max}$. }
%\label{fig.Evsloop}
%\end{figure}
However, in a cycle, it cancels so that the endpoints of the loop (beginning and ending of the loop) become the same, and the energy loop is closed. As we can see in Fig. \ref{fig.PBC_Sys} (d), with increasing system size, the system visits a larger number of minima compared to a small system size. A pertaining question is what happens if the system size is large enough so that it has to visit a nearly infinite number of minima; can it still retrace the path to having a closed loop?}  
\begin{figure*}[htpb]
\includegraphics[scale = 0.28]{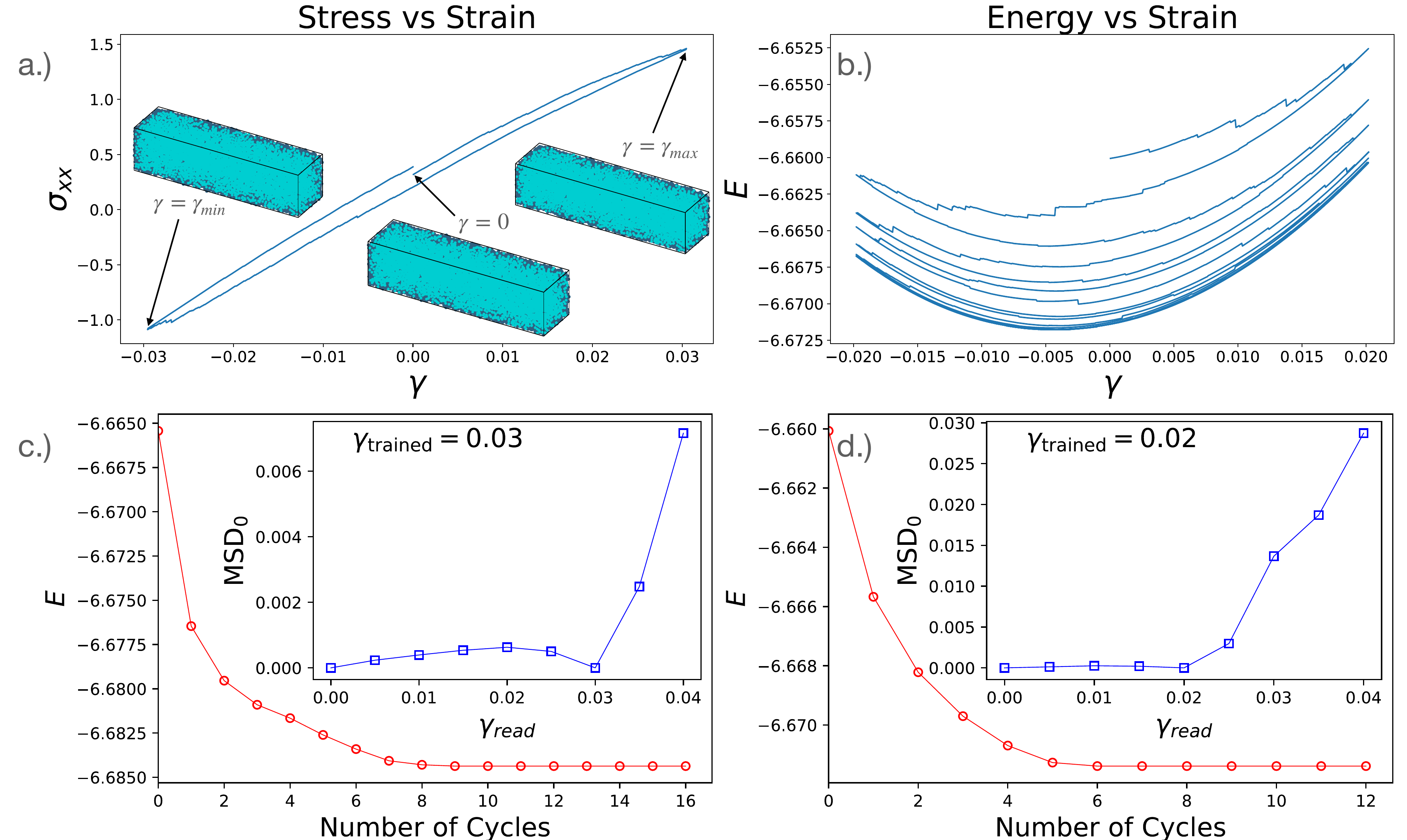}
\caption{Panel (a.) shows the visualisation of one tension-compression cycle applied to a glass nano-rod. Configurations of 0, maximum and minimum strain are also shown. In panel (b.), we see the approach to the limit cycle in a system driven at $\gamma_{max} = 0.02$. Panels (c.) and (d.) show the absorbing state transition for $\gamma_{max} = 0.03$ and $\gamma_{max} = 0.02$, respectively. The insets in each of the panels show the parallel reading protocol. In both cases, the training amplitude can be read off by the sudden large change in the slope of the $MDS_0$ curve. Sequential reading protocols for the same are included in the supplementary material. System size was taken to be $N = 32,000$. 
%\SK{Rishabh: Put all the figures from open surface systems into this figure. You need to put a schematic of the oscillatory compression-decompression process in %anorods and need to put a few more data points in $E_{IS}$ vs number of cycles with different amplitudes. Also include one more memory reading for a smaller %system size, say 5K or so.}
}
\label{fig.absorbingstate}
\end{figure*}

\SK{Interestingly, the number drops in a steady state also increase with system size as a power law with the exponent close to $0.6$ as shown in Fig. \ref{fig.PBC_Sys}(d). Interestingly, the power-law exponent remains close to $0.6$ for all the studied strain amplitude $\gamma_{max}$. It is very tempting to invoke the well-known system size scaling result obtained in uniform shear and oscillatory shear deformation studies on various model systems across spatial dimensions. In steady state, it has been shown that the average strain between two successive plastic drops shows a subextensive scaling behaviour with system size as $\langle\Delta\gamma\rangle \sim N^{-2/3}$ in both two and three-dimensional systems. This suggests that the average strain between two successive plastic drops will vanish in the asymptotic thermodynamic size limit. Now, if we assume the same scaling behaviour for the number of drops in the absorbing state as a function of system size for a given strain amplitude $\gamma_{max}$, then it is natural to expect that the number of plastic drops will grow with system size as $\sim N^{2/3}$ which is very similar to the observation we have. Although one may question the validity of the argument for the absorbing state as $N^{-2/3}$ scaling for the average strain interval between two successive plastic events is true at the steady state. However, similar scaling with a slightly different scaling exponent also holds for the first plastic event statistics \cite{karmakar2010statistical}. Nevertheless, it demonstrates that encoding memory into a large system will be difficult as compared to a smaller system.}
\SK{Surprisingly, we observe that the number of cycles ($\tau$) needed to reach a steady absorbing state for all system sizes and strain amplitudes is universally related to the total number of states the system visits (or equivalently, the total number of plastic drops, $N_{drop}^{tran}$, it encounters) until reaching the absorbing state as $\tau \sim (N_{drop}^{tran})^{\alpha}$, with $\alpha \simeq 0.5$ as shown Fig \ref{fig.PBC_Sys}(f). This observation is indeed very intriguing, but a microscopic understanding of the exponent is still lacking.}

%In Fig. \ref{fig.energyloop} (Right panel), we show the approach to a steady state by investigating the energy measured stroboscopically as a function of number of cycles. We observe that the large system takes long time to reach a steady state. Now question is what happens at $N  \rightarrow \infty$. Will there be a steady state which can encode memory? Currently, we are trying to understand the stability of memory with increasing system size by studying different system sizes, N = 500-100000.         

%\begin{figure}[htp]
%\centering
%\includegraphics[width=0.47\textwidth]{MultiDir/trans_dropvsN_allgamma_corrected}
%\caption{Number of drops to reach the steady state as a function of system size, $N$ for different $\gamma$.  }
%\end{figure}

%\begin{figure}[htp]
%\centering
%\includegraphics[width=0.47\textwidth]{MultiDir/tauvs_transientDrops_allgamma_allN.eps}
%\includegraphics[width=0.47\textwidth]{MultiDir/tauvsTotalDrop_allgamma_corrected.eps}
%\caption{$\tau$ is plotted as a function of total number drops to reach the steady state for all $\gamma$ for different $N$. The line has a slope of around $0.5$. \SK{Monoj: Edit this figure as I told you before with data from individual runs to see whether this works even for individual cases as well. it will be a very important result.} }
%\label{fig.totaldrops}
%\end{figure}

\vskip +0.05in
\noindent{\bf Memory effects in Nanorods: }
\SK{We have just demonstrated that a smaller system size is preferable for encoding memory. However, real-world finite systems inevitably have open boundaries. Thus, it becomes crucial to establish the existence of memory encoding and possible reading protocols for such systems. Here, we take amorphous nano-rods and encode memory by applying tensile and compressive cycles along the long axis. The choice of this particular geometry and protocol serves a 2-fold purpose. Firstly, breaking the symmetry by hand (since the rod has a long axis) ensures that the reading can always be done in the correct direction. This bypasses the need to encode robust memories by training samples in all orthogonal directions, as required for a symmetric cube under oscillatory shear. Thus, this eliminates the risk of inadvertently erasing a memory by applying a reading cycle in the ``wrong" direction, as seen in the inset of Fig. \ref{fig.olddir_mem}. Additionally, using a varied deformation protocol like tension and compression cycles, we show that memory effects in dense amorphous solids aren't confined to just the extensively studied oscillatory shear case. Instead, it highlights the predisposition of such systems to form memories under very general (cyclic) perturbations.}

%\begin{figure}[htp]
%\centering
%\includegraphics[width=0.47\textwidth]{Open_geometry/ensemble1/energy_strain_ens1.png}
%\caption{System is seen to approach a limit cycle. Various plastic drops can be seen as the system anneals to lower energy during the training process.}
%\label{fig.energystrain}
%\end{figure}

\SK{We create the rod geometry by implementing free boundary conditions in the $y$ and $z$ dimensions while taking periodic boundary conditions along $x$ (the long axis of the rod). Each step of the deformation involves scaling the rod length by a factor of $(1 + \epsilon)$ along with providing the corresponding affine displacements to the $x$ coordinates of the particles: $L_x = L_x(1 + \epsilon) $ and $x_i = x_i(1+ \epsilon)$. An energy minimization step follows this.
As before, one cycle involves going from strain: 0 $\rightarrow$ $\gamma_{max}$ $\rightarrow  0 \rightarrow -\gamma_{max} \rightarrow 0$, where $\gamma_{max}$ is the deformation amplitude. This cycle is demonstrated in panel (a.) of Figure \ref{fig.absorbingstate}. Here $\epsilon = 5\times10^{-5}$, and the rod dimensions along the x, y and z axis are taken to be in the ratio of $4:1:1$ respectively. We find that an absorbing state transition occurs in amorphous nanorods, vibrating with an amplitude less than the yielding amplitude $\gamma_{Y}$. This allows for encoding memory of the training amplitude. We read the memory using parallel and sequential reading protocols. Parallel reading is shown in Fig. \ref{fig.absorbingstate} panels (c.) \& (d.),  for $\gamma_{max} = 0.03$ and $0.02$ respectively. The clear kinks in the $MSD_o$ curves in both cases clearly demonstrate the memory encoding and reading in nanorods. Again, the system reaches a diffusive state for amplitude greater than  $\gamma_{Y}$ and rejuvenates.} 
%\begin{figure}[htp]
%\centering
%\includegraphics[width=0.47\textwidth]{Open_geometry/ensemble1/parallel_read_ens_1.png}
%\caption{Parallel reading to read out the training amplitude. The memory is signified by a sharp change in slope of the $MSD_{0}$ curve at $\gamma_{read} = 0.02$.}
%\label{fig.parallelread}
%\end{figure}

%\begin{figure}[htp]
%\centering
%\includegraphics[width=0.47\textwidth]{Open_geometry/ensemble1/sequential_read_ens_1.png}
%\caption{Sequential reading to read out the training amplitude. Again, a sharp signature of the training amplitude at $\gamma_{read} = 0.02$ is visible.}
%\label{fig.seqread}
%\end{figure}

\SK{We see that going from bulk (3D system) to using nanorods (pseudo 1D system) paves the way for a much more efficient way to store and retrieve memory in glassy systems. The efficiency results from using less material, resulting in faster reading and writing time (or faster memory access), a desirable feature in any real-world computational system. Furthermore, less material would, in general, be less dissipative and hence more energy efficient as there are fewer atomic rearrangements (that is, fewer stress drops), which are all considerations to be kept in mind while fabricating new devices. Moreover, a smaller system does not seem to introduce more noise in encoding or reading memory; thus, we feel that our results will encourage future experiments to ascertain the possibility of memory encoding in amorphous nanorods. Degree of annealing \cite{bhaumik2021role,yeh2020glass,sastry2021models,liu2021oscillatory,Maloney2021,mungan2021metastability,parley2021,lamp2022brittle}
is another aspect that might make such systems even better, as our preliminary results suggest faster encodings in better-annealed rods}.

\vskip +0.05in
\noindent{\bf Conclusions:}
\SK{We have shown that the multi-directional shear can be a much better method to encode memory, which will be more stable for reading as one does not have to worry about the shear direction of training beforehand during the reading process and inadvertent memory loss by applying wrong shear protocol during a reading cycle. Our results also show that the large system takes a long time to reach a steady state, which is essential to encode the memory. Thus, reaching a steady state and corresponding memory encoding in a bulk system can be practically impossible. The fundamental relation between the number of shear cycles taken to reach an absorbing state and the total number of states the system visits (equivalently, the number of plastic drops) during the process indicates a deep relation with the underlying complexity of the landscape. It can be an interesting way to study the landscape properties itself. Finally, our results on the existence of absorbing states in amorphous nanorods and demonstration of memory encoding in a few training cycles suggest that amorphous solids at the nanoscale can be an ideal material for encoding memory for industrial applications.}

\vskip +0.05in
\noindent{\bf Acknowledgement:}
\SK{We acknowledge funding by intramural funds at TIFR Hyderabad from the Department of Atomic Energy (DAE) under Project Identification No. RTI 4007. SK acknowledges Swarna Jayanti Fellowship grants DST/SJF/PSA01/2018-19 and SB/SFJ/2019-20/05 from the Science and Engineering Research Board (SERB) and Department of Science and Technology (DST) and the National Super Computing Mission (NSM) grant $\mathrm{DST/NSM/R\&D\_HPC\_Applications/2021/29}$ for generous funding. Most of the computations are done using the HPC clusters procured using Swarna Jayanti Fellowship grants DST/SJF/PSA01/2018-19, SB/SFJ/2019-20/05 and Core Research Grant CRG/2019/005373. MA acknowledges support from NSM grant $\mathrm{DST/NSM/R\&D\_HPC\_Applications/2021/29}$ for financial support.}
\bibliographystyle{apsrev4-1}
\bibliography{MemoryStability}

\end{document}

%\begin{titlepage}
\title{Encoding Robust and Fast Memories in Bulk and Nanoscale Amorphous Solids}

\author{Monoj Adhikari}
\thanks{Equal contributions}
\affiliation{Tata Institute of Fundamental Research, 36/P, Gopanpally Village, Serilingampally Mandal, Ranga Reddy District, Hyderabad 500046, Telangana, India}
\author{Rishabh Sharma}
\thanks{Equal contributions}
\affiliation{Tata Institute of Fundamental Research, 36/P, Gopanpally Village, Serilingampally Mandal, Ranga Reddy District, Hyderabad 500046, Telangana, India}
\author{Smarajit Karmakar}
\email[Corresponding author: ]{smarajit@tifrh.res.in}
\affiliation{Tata Institute of Fundamental Research, 36/P, Gopanpally Village, Serilingampally Mandal, Ranga Reddy District, Hyderabad 500046, Telangana, India}

\maketitle

%\section{Energy and number of drops}
%\begin{figure}[htp]
%\centering
%\includegraphics[width= 0.48\textwidth]{MultiDir/EvsCycle_gamma06.eps}
%\includegraphics[width= 0.48\textwidth]{MultiDir/Compare_N5vs50k.eps}
%\includegraphics[width= 0.48\textwidth]{MultiDir/gamma06tauvsN.eps}
%\caption{ The system is trained with $0.06$. Left:$E(\gamma=0)$ is plotted as a function of $N_{cyclces}$ in the steady state for two different system sizes as shown the legend. Right: $\Delta E$ is plotted as a function of $\gamma$ in the steady state for two different system sizes as shown the legend. $\Delta E = E_{\gamma}-E_{fit}$, where $E_{\gamma}$ is the actual energy in the steady state in a cycle and $E_{fit}$ is obtained by parabolic fitting of the $E_{\gamma}$. }
%\label{SMfig.energyloop}
%\end{figure}

%\section{steady state drops}
%\begin{figure}[htp]
%\centering
%\includegraphics[width= 0.48\textwidth]{MultiDir/tauvs_N_allgamma.eps}
%\includegraphics[width=0.48\textwidth]{MultiDir/NumberofDrop_N_allgamma.eps}
%\caption{The system is trained with $\gamma=0.02$, $\gamma=0.04$, $\gamma=0.06$.  Bottom: The number of drops are plotted as a function of system sizes at the steady state. Number of drops also show a power law and the slope is same for every $\gamma$.}
%\label{SMfig.steadysatteDrop}
%\end{figure}

\noindent{\bf{Details of BMLJ model}}
The interaction potential, with a quadratic cut-off, is given by 
\begin{eqnarray}
V_{\alpha \beta}(r) &=& 4\epsilon _{\alpha\beta}\left[ \left(\frac{\sigma_{\alpha \beta}}{r}\right)^{12}-\left(\frac{\sigma_{\alpha \beta}}{r}\right)^{6}\right. \nonumber \\ \nonumber
&+& \left. c_0+c_2\left(\frac{r}{\sigma_{\alpha \beta}}\right)^2\right], r_{\alpha \beta} \leq r_{c, \alpha \beta } \\
 &=&0 ,\hspace{1.94cm} r_{\alpha \beta } > r_{c, \alpha \beta } 
\end{eqnarray}
where $\alpha ,\beta$ $\in$ (A,B),  $\epsilon_{AB}/\epsilon_{AA}  = \epsilon_{BA}/\epsilon_{AA} = 1.5$, $\epsilon_{BB}/$ $\epsilon_{AA}  = 0.5 $, and 
$\sigma_{AB}/\sigma_{AA} = \sigma_{BA}/\sigma_{AA} = 0.8 $, $\sigma_{BB}/\sigma_{AA} = 0.88 $. The interaction potential has cut off, $r_{c, \alpha \beta }$ = $2.5\sigma_{\alpha \beta}$.  We present results in reduced units, with units of length, energy and time scales being $\sigma_{AA}$, $\epsilon_{AA}$  and  $\sqrt \frac{\sigma^2_{AA}m_{AA}}{\epsilon_{AA}} $ respectively. 

\vskip +0.05in
\noindent{\bf Preperation Protocol for nano rods:}
The glass rods in this study were formed by taking the KA mixture at a density of $\rho = 1.2$ and equilibriating it at a high temperature of $1.0$. This liquid was then cooled down to a low temperature of 0.01 using a cooling rate of $\dot{T}= 10^{-1}$ in the LJ time units. This  was followed by an annealing phase assisted by active dynamics, using the protocol demonstrated in \cite{sharma2023activeannealing} with an active forcing $f_0 = 1.9$. Following this the system was run in an NPT ensemble for additional 1000 MD steps at zero pressure. This was ultimately followed by removing periodic boundaries in the Y and Z directions, and minimizing the energy using conjugate gradient algorithm. Nos\'e-Hoover thermo and barostats were used. The time step for integration was taken to be 0.005. 

\begin{figure*}[htpb]
\centering
\includegraphics[scale = 0.25]{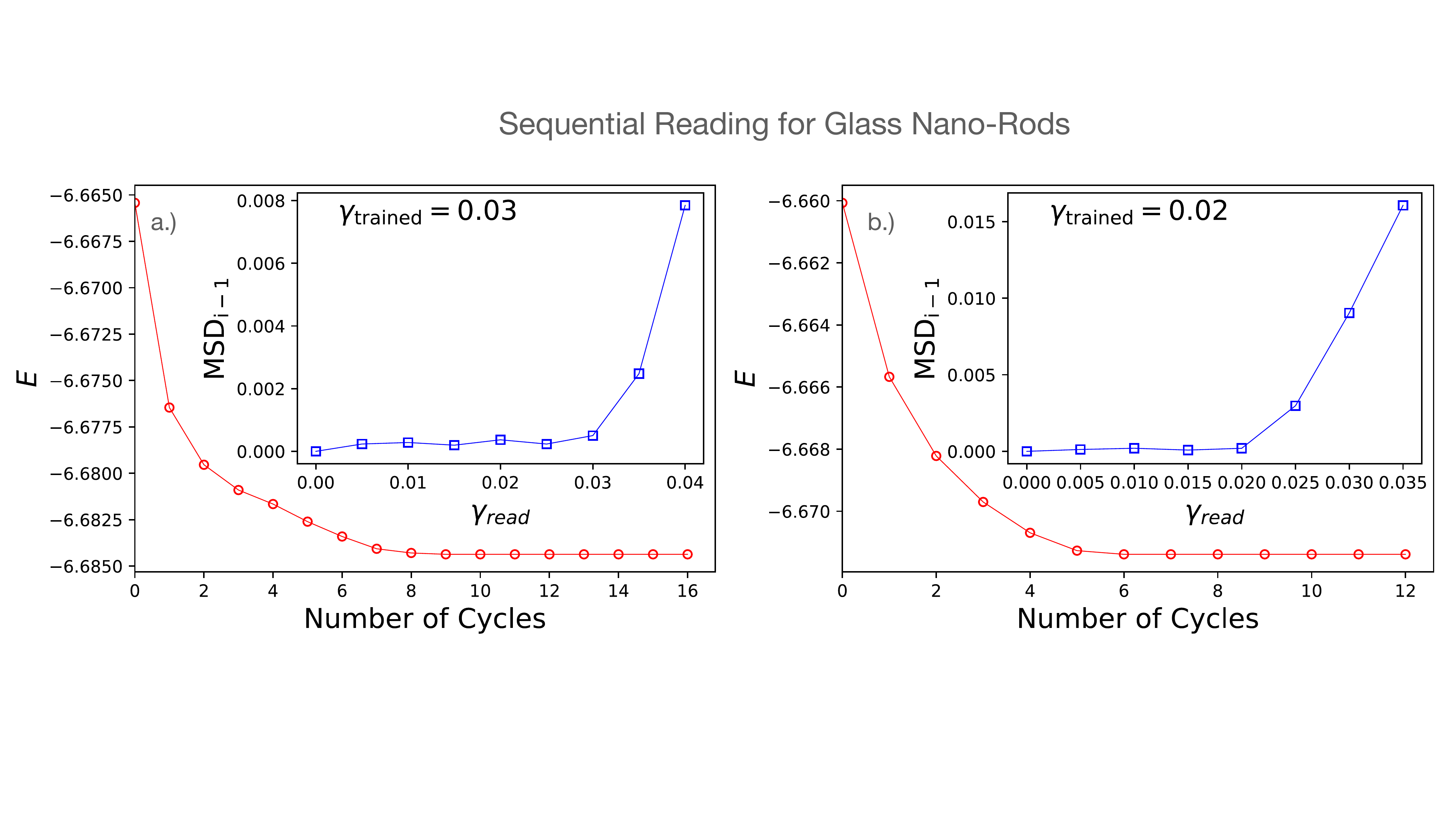}
\caption{ Sequential reading protocol for $\gamma_{max} = 0.03$ and $\gamma_{max} = 0.02$ are shown in panels (a.) and (b.) respectively.
Here cycles of increasing strain amplitudes are applied to the same sample and displacement after each cycle is calculated with respect to the final configuration of the previous cycle.}
\label{fig.absorbingstate}
\end{figure*}

\bibliographystyle{apsrev4-1}
\bibliography{MemoryStability}